\documentclass{elsarticle}
\usepackage{graphicx}%
\usepackage{multirow}%
\usepackage{amsmath,amssymb,amsfonts}%
\usepackage{amsthm}%
\usepackage{mathrsfs}%
\usepackage[title]{appendix}%
\usepackage{xcolor}%
\usepackage{textcomp}%
\usepackage{manyfoot}%
\usepackage{booktabs}%
\usepackage{algorithm}%
\usepackage{algorithmicx}%
\usepackage{algpseudocode}%
\usepackage{listings}%
\usepackage{verbatim}
\usepackage{enumitem}
 \usepackage{comment}
 \usepackage{amsmath}
 \usepackage{booktabs}
\usepackage{amssymb}
\usepackage{url}

\journal{}
\begin{document}

\begin{frontmatter}

\title{Large Deviations and Critical Scaling in the Finite-Population Granovetter Threshold Model}

\author[mymainaddress]{Jos\'e F.  Fontanari}

\address[mymainaddress]{Instituto de F\'{\i}sica de S\~ao Carlos,  Universidade de S\~ao Paulo,  13566-590 S\~ao Carlos,  S\~ao Paulo,  Brazil}

\begin{abstract}
We develop a systematic asymptotic analysis of the exact finite-population solution of Granovetter's threshold model, in which a single initial instigator triggers a cascade among agents with independently distributed adoption thresholds. By expressing the propagation probability as an order-statistic boundary-crossing event, we formulate its macroscopic large-deviation rate as a constrained relative-entropy minimization problem. For Beta-distributed thresholds, we obtain explicit solutions for all shape parameters $\alpha$ and $\beta$, identifying four possible optimal empirical threshold profiles: fully free, fully pinned, pinned--free, and free--pinned. These profiles determine both the exponential probability of macroscopic cascades and the threshold configurations responsible for their occurrence. Importance-sampling estimates confirm the analytical rate functions, while sequential Monte Carlo simulations independently reproduce the predicted optimal profiles.  We also revisit the critical scaling near $\alpha_c=1$ and confirm the principal finite-size scaling laws: the mean active fraction decays as $N^{-1/2}$ for uniform thresholds and as $N^{-1}$ for $\alpha=1$ and $0<\beta<1$. In the latter regime, we further derive an explicit scaling function for the logarithmic critical window from the microscopic branching-process limit.  Our results provide a general framework for determining the probabilities and characteristic configurations of rare collective events in threshold-driven cascade processes.
\end{abstract}


\end{frontmatter}

\section{Introduction}\label{sec:intro}

Collective actions often begin with a small number of individuals whose decisions encourage others to follow~\cite{granovetter1978,schelling1978}. A protest may grow because each new participant persuades additional people to join~\cite{granovetter1978,kuran1989}; a new technology may spread as its adoption becomes increasingly visible~\cite{bass1969,young2009}; and a financial panic may develop when the actions of a few investors induce others to withdraw their funds~\cite{diamond1983,bikhchandani1992}. A central question in all these examples is whether a small initial disturbance will disappear or develop into a large-scale collective event~\cite{watts2002,castellano2009,jusup2022}.

Threshold models provide a simple framework for studying this question. In Granovetter's model~\cite{granovetter1978}, each individual has a threshold representing the minimum fraction of the population that must already be participating before that individual is willing to join. These thresholds differ across individuals and are drawn from a common probability distribution. As participation increases, more thresholds may be exceeded, allowing a cascade to propagate through the population.

For a very large population, the evolution of the active fraction is usually described by deterministic equations involving the cumulative threshold distribution $F(x)$~\cite{granovetter1983}. Such equations are useful for identifying stable collective states and the conditions under which a large cascade can develop. However, they do not fully describe what happens in a finite population. Even when the deterministic description predicts that a small disturbance cannot grow, a particular realization of the individual thresholds may be unusually favorable to propagation. A cascade initiated by a single individual can then recruit a substantial fraction of the population, although the probability of such an event may be extremely small.

Understanding these exceptional cascades is the subject of the present work. Rather than asking whether a macroscopic cascade is typical, we ask how unlikely it is when the population size $N$ becomes large. More precisely, we seek the probability that a single permanent instigator produces a cascade involving a finite fraction $x$ of the population, even when this outcome requires an atypical realization of the thresholds. Such probabilities are naturally described by large-deviation theory, which characterizes events whose probabilities decrease exponentially with system size~\cite{touchette2009}.

An exact finite-population formulation of this problem was obtained recently~\cite{fontanari2026}. Its central observation is that successful propagation imposes a sequence of conditions on the thresholds of the recruited individuals. If their thresholds are arranged in increasing order, $X_{(1)}\le X_{(2)}\le\cdots$, a cascade can recruit $k-1$ individuals in addition to the initial instigator only if
\begin{equation}
X_{(j)}\le \frac{j}{N},
\qquad j=1,\ldots,k-1.
\label{eq:intro_constraints}
\end{equation}
The meaning of these inequalities is straightforward. The first recruit must be willing to join when only one individual is active; the second must be willing to join when two individuals are active; and so on. If any one of these conditions fails, the cascade cannot continue through the proposed group of recruits.

This sequential requirement is the main difficulty of the problem. It is not sufficient for the population to contain many individuals with relatively low thresholds. Those thresholds must also be distributed appropriately so that the cascade can advance through every intermediate stage. Consequently, estimating the probability of a large cascade requires more than counting the individuals whose thresholds lie below the final active fraction.

For uniformly distributed thresholds, the probability of satisfying Eq.~(\ref{eq:intro_constraints}) is closely related to the classical parking-function problem~\cite{konheim1966,stanley1999,pitman2006}, for which exact counting formulas are available. For general threshold distributions, the corresponding probability is considerably more difficult to evaluate. In particular, a conventional saddle-point approximation may fail because the most probable threshold configuration producing a large cascade need not lie in the interior of the allowed region. This difficulty was highlighted in a recent Comment on the finite-population formulation~\cite{zeraoulia_comment}.

The physical origin of this difficulty becomes clearer when we examine the threshold distribution among the individuals participating in a rare cascade. Suppose that the cascade reaches a fraction $x=k/N$ of the population. In the large-$N$ limit, the sequential conditions in Eq.~(\ref{eq:intro_constraints}) require the empirical cumulative distribution $G(t)$ of the recruited thresholds to satisfy
\begin{equation}
G(t)\ge \frac{t}{x},
\qquad 0\le t\le x.
\label{eq:intro_barrier}
\end{equation}
Thus, the empirical distribution cannot fall below a straight-line boundary. If it did, there would be some intermediate stage at which too few individuals were willing to participate, and propagation would stop.

The central idea of our approach is to identify the most probable threshold configuration that sustains a macroscopic cascade. We formulate this problem as a constrained relative-entropy minimization, which determines both the leading exponential decay rate of the propagation probability and the typical threshold configuration associated with a rare cascade. Importantly, the continuum variational problem must be connected to the exact finite-population constraints. We establish this connection through a change of probability measure, quantile coupling, and the exact enumeration of uniform parking functions.

We apply this framework to the two-parameter family of Beta threshold distributions, $\mathrm{Beta}(\alpha,\beta)$, which encompasses decreasing, increasing, bell-shaped, and U-shaped densities. For all $\alpha,\beta>0$, we obtain explicit expressions for the macroscopic propagation rate and identify the corresponding optimal threshold profiles. The analytical predictions are verified numerically using importance sampling and sequential Monte Carlo simulations, which independently reproduce the predicted threshold configurations.

We also revisit the critical scaling near $\alpha_c=1$, previously investigated in Ref.~\cite{fontanari2026}. We confirm the principal finite-size scaling laws for the mean final active fraction and derive an explicit scaling function for the logarithmic critical window when $0<\beta<1$. This result connects the finite-size behavior to the microscopic branching-process description and complements the macroscopic large-deviation analysis.

The paper is organized as follows. Section~\ref{sec:exact} presents the exact finite-population formulation, and Section~\ref{sec:entropy} develops the constrained-entropy method and establishes its asymptotic validity. Section~\ref{sec:beta} derives the propagation rates and optimal profiles for the Beta family, while Section~\ref{sec:numerics} presents their numerical verification. Section~\ref{sec:critical-scaling} investigates the critical finite-size scaling, and Section~\ref{sec:discussion} concludes the paper.


\section{Exact finite-population propagation event}
\label{sec:exact}

We consider a population of $N$ agents, one of whom is designated as a permanent instigator. This individual participates unconditionally and initiates the cascade. The remaining $N-1$ agents have independent activation thresholds $X_1,\ldots,X_{N-1}$, drawn from a continuous cumulative distribution $F(t)$ with probability density $f(t)$ supported on $[0,1]$. An inactive agent becomes active whenever the fraction of active individuals reaches or exceeds its threshold. Once activated, an agent remains active, so that the cascade proceed irreversibly until no further recruitment is possible~\cite{granovetter1978}.

Our first objective is to determine the probability $P_N(k)$ that the cascade stops with exactly $k$ active agents, including the initial instigator. As shown in Ref.~\cite{fontanari2026}, this probability can be expressed exactly in terms of the thresholds of the individuals who eventually participate.

To see how this works, suppose that the cascade terminates with $k$ active agents. Besides the instigator, there must be a group of $k-1$ individuals
whose thresholds allow them to become active. There are $\binom{N-1}{k-1}$ possible choices for this group.

Consider one such group and arrange its thresholds in increasing order, $X_{(1)}\le X_{(2)}\le\cdots\le X_{(k-1)}$. Initially, only the instigator is active, so the first recruit must have a threshold not exceeding $1/N$. Once this individual joins, the active fraction becomes $2/N$, allowing the second recruit to participate if its threshold does not exceed $2/N$. The same reasoning applies at every subsequent stage.

Consequently, the group can be recruited completely if and only if its ordered thresholds satisfy $X_{(j)}\le j/N$ for every $j=1,\ldots,k-1$, as stated in Eq.~(\ref{eq:intro_constraints}). These inequalities express the sequential nature of the cascade: at each stage, there must be at least one additional individual willing to join; otherwise, recruitment stops.

We denote the probability that a specified group of $k-1$ individuals satisfies all these conditions by
\begin{equation}
Q_{N,k}
=
\mathbb{P}\left\{
X_{(j)}\le\frac{j}{N},
\quad j=1,\ldots,k-1
\right\},
\label{eq:Q}
\end{equation}
with $Q_{N,1}=1$. We refer to $Q_{N,k}$ as the \emph{propagation probability}.

Successful recruitment of the selected group is not sufficient for the cascade to stop at size $k$. We must also require that none of the remaining
$N-k$ agents becomes active. Since the final active fraction is $k/N$, this happens precisely when all their thresholds exceed $k/N$. The probability of this event is $[1-F(k/N)]^{N-k}$.

Because the thresholds of distinct agents are independent, the probabilities of recruitment and arrest can be multiplied. Moreover, two different groups of $k-1$ recruits cannot both represent the final active population: an agent outside one such group has threshold greater than $k/N$ and therefore cannot belong to another successfully recruited group.

Summing over the possible choices of recruits gives the exact cascade-size distribution
\cite{fontanari2026},
\begin{equation}
P_N(k)
=
\binom{N-1}{k-1}
Q_{N,k}
\left[1-F\left(\frac{k}{N}\right)\right]^{N-k},
\label{eq:P}
\end{equation}
for $k=1,\ldots,N$. As usual, the factor associated with the absence of inactive agents is understood to be unity when $k=N$ (with $[1-F(1)]^0 \equiv 1$), so that $P_N(N) = Q_{N,N}$.

Equation~(\ref{eq:P}) separates the probability of a cascade of size $k$ into three contributions: the number of possible recruited groups, the
probability that a chosen group can be activated, and the probability that recruitment stops once this group has become active. The only nontrivial quantity is the propagation factor $Q_{N,k}$, because it involves several simultaneous conditions on ordered random thresholds.

\subsection{Exact recursion for the propagation probability}

The propagation probability $Q_{N,k}$ can be computed recursively for any finite $N$~\cite{fontanari2026}. To obtain the recursion, consider the event that all $k-1$ thresholds lie below $(k-1)/N$, and subtract the mutually exclusive events in which recruitment stops at an earlier size $j<k$. This gives
\begin{equation}
Q_{N,k}
=
F\left(\frac{k-1}{N}\right)^{k-1} -
\sum_{j=1}^{k-2}
\binom{k-1}{j-1}Q_{N,j}
\left[
F\left(\frac{k-1}{N}\right)
-
F\left(\frac{j}{N}\right)
\right]^{k-j},
\label{eq:Q-rec}
\end{equation}
for $k\ge2$, with $Q_{N,1}=1$. This recursion is equivalent to the normalization identity derived in Ref.~\cite{fontanari2026}, but avoids subtracting small probabilities from unity and is therefore numerically preferable at moderate $N$. Nevertheless, for macroscopic cascades, $Q_{N,k}$ may be exponentially smaller than the terms entering the recursion, leading to severe cancellation errors. Moreover, the recursion does not directly reveal the
large-deviation rate, which motivates the asymptotic approach developed below.

\subsection{Separating the two requirements for propagation}

We now consider macroscopic cascades, for which the final active fraction approaches a fixed value $x$ as the population grows:
\begin{equation}
\frac{k}{N}\longrightarrow x,
\qquad 0<x<1.
\end{equation}
Our aim is to determine how rapidly the propagation probability decreases with $N$.

A useful first step is to separate two different requirements for successful recruitment. First, all $k-1$ recruited individuals must have sufficiently low
thresholds. Second, these thresholds must be distributed appropriately to allow recruitment to proceed without interruption.

To make this separation explicit, define
\begin{equation}
n=k-1,
\qquad
a=\frac{n}{N}.
\end{equation}
Here $n$ counts the recruited agents, excluding the permanent instigator, while $a$ is the largest threshold compatible with their complete recruitment.

The final inequality in Eq.~(\ref{eq:intro_constraints}) requires the largest recruited threshold to be at most $a$. Therefore, all $n$ thresholds must lie in the interval $(0,a)$. Because the thresholds are independent, the probability that they all lie in this interval is simply $F(a)^n$.

Now suppose that we already know that all $n$ thresholds belong to $(0,a)$. Their conditional probability density is
\begin{equation}
q_a(t)
=
\frac{f(t)}{F(a)}
\mathbf{1}_{(0,a)}(t).
\label{eq:qa}
\end{equation}
In other words, $q_a$ is the original threshold distribution restricted to $(0,a)$ and normalized to unit probability.
The propagation probability can then be written exactly as
\begin{equation}
Q_{N,k}=F(a)^nR_{N,k},
\label{eq:factor}
\end{equation}
where $R_{N,k}$ is the conditional probability that the $n$ thresholds, now independently drawn from $q_a$, satisfy the sequential recruitment conditions.
Since $a=n/N$, these conditions can be written as
\begin{equation}
E_n=
\left\{
X_{(j)}\le\frac{aj}{n},
\quad j=1,\ldots,n
\right\},
\label{eq:E}
\end{equation}
so that
\begin{equation}
R_{N,k}=\mathbb{P}_{q_a}(E_n).
\end{equation}
Notice that $R_{N,k}$ is labeled by the final cascade size $k$, whereas $E_n$ involves only the $n=k-1$ recruited agents. The two quantities therefore describe the same propagation process from different perspectives.

The factorization~(\ref{eq:factor}) is important because it separates the overall scarcity of sufficiently low thresholds, represented by $F(a)^n$, from the additional difficulty of arranging these thresholds in a way that sustains the cascade, represented by $R_{N,k}$.
As we shall see, the second factor can be as important as the first. In particular, even after conditioning all thresholds to lie below $a$, successful propagation may remain exponentially unlikely.

\subsection{The uniform case and parking functions}
\label{subsec:parking}

When the conditional density $q_a$ is uniform on $(0,a)$, the rescaled thresholds $U_i=X_i/a$ are independent and uniform on $[0,1]$. The propagation event $E_n$ becomes
\begin{equation}
U_{(j)}\le\frac{j}{n},
\qquad j=1,\ldots,n.
\label{eq:uniform_event}
\end{equation}
These inequalities are equivalent to the classical parking-function conditions~\cite{konheim1966,stanley1999,pitman2006}. Indeed, setting $p_i=\lceil nU_i\rceil$ maps the event onto the successful parking of $n$ cars with preferences $p_i\in\{1,\ldots,n\}$. Since exactly $(n+1)^{n-1}$ of the $n^n$
preference sequences are successful, we obtain
\begin{equation}
\mathbb{P}_{\mathrm{unif}}(E_n)
=
\frac{(n+1)^{n-1}}{n^n}
\sim\frac{e}{n}.
\label{eq:parking}
\end{equation}
Thus, the sequential propagation constraints introduce only an algebraic, rather than exponential, probability cost under uniform sampling.

For an originally uniform threshold distribution, $F(t)=t$, the factorization~(\ref{eq:factor}) therefore gives the exact result
\begin{equation}
Q_{N,k}
=
\frac{k^{k-2}}{N^{k-1}},
\label{eq:Q-unif}
\end{equation}
in agreement with Ref.~\cite{fontanari2026}.

For nonuniform threshold distributions, however, the conditional propagation probability $R_{N,k}$ may decay exponentially with $N$. Determining
this additional exponential cost is the large-deviation problem addressed next.

\section{Constrained entropy and a finite-sample rate criterion}
\label{sec:entropy}

In the preceding section, we separated the propagation probability into two factors, $Q_{N,k}=F(a)^nR_{N,k}$, with $n=k-1$ and $a=n/N$. The first factor is the probability that all recruited individuals have thresholds below $a$. The second is the probability that these thresholds are sufficiently well distributed to sustain recruitment at every stage. We are interested in macroscopic cascades, for which $k/N\to x\in(0,1)$ as $N\to\infty$. Since $n/N\to x$ and $a\to x$, the contribution of the first factor to the exponential decay rate follows immediately:
\begin{equation}
\lim_{N\to\infty}\frac{1}{N}\log F(a)^n
=x\log F(x).
\end{equation}
The main difficulty is therefore to determine the large-$N$ behavior of $R_{N,k}$. This is an example of a general problem in statistical physics: how can we calculate the probability that many random variables satisfy a large number of simultaneous constraints, especially when the required event is exponentially unlikely?

We address this question in two steps. First, we identify the most probable configuration satisfying the constraints by minimizing a relative-entropy cost. We then show how a suitable change of probability measure connects this continuum prediction to the exact finite-sample problem.

\subsection{The most probable realization of a rare event}
\label{subsec:variational}

Consider $n$ independent thresholds drawn from the conditioned density $q_a(t)$ introduced in Eq.~(\ref{eq:qa}). For large $n$, their empirical distribution will ordinarily be close to $q_a$. This is a consequence of the law of large numbers: the fraction of sampled thresholds belonging to any given interval approaches the probability assigned to that interval by $q_a$.

A rare event, however, may require an empirical distribution very different from the one that would normally be observed. The central idea of large-deviation theory is that not all such departures are equally unlikely. Among the distributions compatible with the rare event, the one requiring the smallest statistical departure from the original distribution determines its leading exponential probability.

To make this statement quantitative, let $g(t)$ be a normalized probability density on $[0,x]$, and denote its cumulative distribution by
\begin{equation}
G(t)=\int_0^t g(s),ds,
\end{equation}
with $G(0)=0$ and $G(x)=1$.

The interpretation of $G(t)$ is straightforward: it represents the fraction of the recruited individuals whose thresholds do not exceed $t$. Since the recruits constitute a fraction approaching $x$ of the entire population, the corresponding fraction of the population is $xG(t)$ in the macroscopic limit. The single permanent instigator makes a contribution of order $1/N$, which vanishes in this limit.

For recruitment to proceed, the population must contain sufficiently many individuals with low thresholds at every intermediate stage. In particular, the fraction $xG(t)$ of individuals whose thresholds do not exceed $t$ must be at least $t$. Otherwise, the cascade cannot reach the active fraction $t$ and subsequently recruit individuals with higher thresholds. The resulting continuum propagation constraint is
\begin{equation}
G(t)\ge\frac{t}{x},
\qquad 0\le t\le x.
\label{eq:cascade_constraint}
\end{equation}
It is instructive to see how this condition emerges from the exact finite-population requirements. Recall from Eq.~(\ref{eq:E}) that successful recruitment
of $n=k-1$ individuals requires
\begin{equation}
X_{(j)}\le\frac{aj}{n},
\qquad j=1,\ldots,n,
\end{equation}
where $a=n/N$. Introduce the empirical cumulative distribution of the recruited thresholds,
\begin{equation}
\widehat G_n(t)
=
\frac1n\sum_{i=1}^n
\mathbf{1}_{\{X_i\le t\}}.
\end{equation}
The order-statistic inequalities are exactly equivalent to
\begin{equation}
\widehat G_n\left(\frac{aj}{n}\right)
\ge\frac{j}{n},
\qquad j=1,\ldots,n.
\end{equation}
Thus, at each of the $n$ discrete recruitment stages, the empirical cumulative distribution must reach or exceed the corresponding point on the diagonal. As $N\to\infty$ with $n/N\to x$, these grid points become dense in $[0,x]$, leading to the continuum constraint~(\ref{eq:cascade_constraint}).

Geometrically, this constraint requires the cumulative distribution $G(t)$ to lie on or above the straight line connecting $(0,0)$ to $(x,1)$. The line itself corresponds to the uniform density $g(t)=1/x$, for which the supply of individuals with sufficiently low thresholds is exactly at the minimum required by the continuum constraint. A distribution lying strictly above this line has an excess of low-threshold individuals at the corresponding values of $t$. Conversely, if $G(t^*)<t^*/x$ at some intermediate point $t^*\in(0,x)$, there are too few individuals with thresholds below $t^*$ to sustain recruitment up to that point. The cascade must therefore stop before reaching its intended macroscopic size. The boundary $G(t)=t/x$ is consequently a marginal propagation condition. As we shall see, the most probable distribution compatible with successful propagation may coincide with this boundary over some intervals while remaining strictly above it elsewhere.

We denote by $\mathcal{A}_x$ the set of normalized densities satisfying the continuum propagation constraint:
\begin{equation}
\label{eq:A}
\mathcal{A}_x=
\left\{
g\ge0:
\int_0^x g(t)\,dt=1,\quad
\int_0^t g(s)\,ds\ge\frac{t}{x}
\ \text{for all }t\in[0,x]
\right\}.
\end{equation}
How unlikely is it for independent samples from $q_x$ to produce an empirical distribution close to a prescribed profile $g$?

Sanov's theorem provides the answer at the exponential level~\cite{dembo1998,touchette2009}. For large $n$, the probability of observing an empirical distribution near $g$ has leading exponential
behavior
\begin{equation}
\mathbb{P}\{\text{empirical distribution near }g\}
\asymp
\exp[-nD_{\mathrm{KL}}(g\Vert q_x)],
\end{equation}
where
\begin{equation}
\label{eq:I}
D_{\mathrm{KL}}(g\Vert q_x)
=
\int_0^x
g(t)\log\frac{g(t)}{q_x(t)}\,dt
\end{equation}
is the Kullback--Leibler divergence, also called relative entropy. Here $\asymp$ denotes equality at the leading exponential level, without specifying algebraic prefactors. The relative entropy measures the
statistical cost of replacing the original distribution $q_x$ by an atypical distribution $g$. It is nonnegative and vanishes only when the two distributions coincide almost everywhere. Thus, the typical distribution has zero cost, while atypical distributions have positive costs.

For an event that can be realized by many different empirical profiles, the dominant contribution comes from the admissible profile with the smallest relative entropy. This is the large-deviation counterpart of the saddle-point principle familiar from statistical mechanics. We are therefore led to the constrained minimization problem
\begin{equation}
\mathcal{I}_F(x)
=
\inf_{g\in\mathcal{A}_x}
D_{\mathrm{KL}}(g\Vert q_x).
\label{eq:entropy-minimum}
\end{equation}
The minimizing density, denoted by $g_x^*$, describes the most probable threshold distribution among realizations sustaining a macroscopic cascade. The continuum argument suggests that
\begin{equation}
R_{N,k}\asymp
\exp[-n\mathcal{I}_F(x)].
\end{equation}
Since $n/N\to x$, combining this expression with the factor $F(a)^n$ leads to the candidate propagation rate
\begin{equation}
\psi_Q(x)
=
x\log F(x)-x\mathcal{I}_F(x).
\label{eq:psi-candidate}
\end{equation}

There is another useful interpretation of this formula. Since $q_x(t)=f(t)/F(x)$ on $[0,x]$, we have
\begin{equation}\label{eq:KL_qg}
D_{\mathrm{KL}}(g\Vert q_x)
=
\int_0^x
g(t)\log\frac{g(t)}{f(t)}\,dt
+\log F(x).
\end{equation}
Consequently,
\begin{equation}
\psi_Q(x)
=
-x\inf_{g\in\mathcal{A}_x}
\int_0^x
g(t)\log\frac{g(t)}{f(t)}\,dt.
\end{equation}
Because every admissible density $g$ is normalized on $[0,x]$ and vanishes outside this interval, the integral is precisely its relative entropy with respect to the original threshold density $f$.

The propagation rate is therefore determined by the least costly statistical rearrangement of the original threshold distribution that satisfies the propagation barrier, Eq.~(\ref{eq:cascade_constraint}). The minimizing profile need not resemble the original distribution. In particular, its cumulative distribution may coincide with the propagation boundary over a finite interval, corresponding to a region where the constraint is saturated. Such boundary contacts will play an important role in the following sections.

\subsection{Why the continuum argument is not sufficient}
\label{subsec:discrete-gap}

The entropy minimization provides a natural prediction for the exponential decay rate. However, it does not automatically establish that the exact finite-population probability has this rate. The reason is that the original event $E_n$ is more restrictive than the continuum description might suggest. It requires every one of the $n$ ordered thresholds to satisfy its corresponding inequality, without allowing even a small violation. Three difficulties deserve particular attention.

First, standard large-deviation arguments describe probabilities of sets of empirical distributions. A relaxed event that permits small violations of the constraint need not have the same lower bound as the exact event. Second, the optimal cumulative distribution may coincide with the boundary $G(t)=t/x$ over an entire interval. On such a pinned interval, ordinary sample fluctuations occur on both sides of the boundary. Requiring the empirical distribution to remain on the allowed side at every point may introduce an additional probability cost. Third, the earliest recruitment steps involve only a few individuals. These microscopic constraints remain present no matter how large the population becomes. A continuum description alone does not guarantee that recruitment can pass through these initial bottlenecks.

The key question is whether these finite-sample restrictions modify only the subexponential factors or also the leading exponential decay rate. We will show that, for the Beta distributions considered here, they do not change the exponential rate predicted by entropy minimization. The proof combines a change of probability measure, closely related to importance sampling in statistical mechanics, with the exact parking-function identity (\ref{eq:parking}).

\subsection{Changing the sampling distribution}
\label{subsec:tilt}

The difficulty of estimating a rare probability is often caused by the distribution from which samples are generated. Under the conditioned density $q_a$, successful propagation may be exponentially unlikely. We can instead generate thresholds from another distribution under which successful propagation is more common, and correct for the change in sampling probabilities. This is the principle of importance sampling~\cite{Hammersley1964}. Closely related ideas appear in umbrella sampling and histogram reweighting in statistical mechanics~\cite{Ferrenberg1988,Kumar1992}.

Let $h_a(t)$ be an auxiliary probability density on $(0,a)$, with cumulative distribution
\begin{equation}
H_a(t)=\int_0^t h_a(s)\,ds.
\end{equation}
We will eventually choose $h_a$ to be the entropy-minimizing density $g_a^*$ obtained from the continuum problem, evaluated at the finite endpoint $a=n/N$. For the change of measure to be valid, we require $h_a(t)>0$ wherever $q_a(t)>0$, apart from sets of zero measure. Define the logarithmic likelihood ratio
\begin{equation}
\ell_a(t)=
\log\frac{q_a(t)}{h_a(t)}.
\end{equation}
A configuration of thresholds $Y_1,\ldots,Y_n$ generated from $h_a$ receives the statistical weight
\begin{equation}
W_n=
\prod_{i=1}^n
\frac{q_a(Y_i)}{h_a(Y_i)}
=
\exp\left[
\sum_{i=1}^n\ell_a(Y_i)
\right].
\end{equation}
The weight compensates exactly for the use of the auxiliary distribution. Therefore, the conditional propagation probability can be expressed as
\begin{equation}
R_{N,k}
=
\mathbb{E}_{h_a}
\left[
\mathbf{1}_{E_n}
\exp\left(
\sum_{i=1}^n\ell_a(Y_i)
\right)
\right].
\label{eq:rn-identity}
\end{equation}
Here $Y_1,\ldots,Y_n$ are independent samples from the auxiliary density $h_a$. Importantly, the propagation event $E_n$ is exactly the same as under the original density $q_a$:
\begin{equation}
E_n=
\left\{
Y_{(j)}\le\frac{aj}{n},
\quad j=1,\ldots,n
\right\}.
\end{equation}
Thus, the indicator $\mathbf{1}_{E_n}$ equals one when all these inequalities are satisfied and zero otherwise. The change of measure modifies only the probability assigned to each threshold configuration, not the conditions defining successful propagation. The likelihood weight compensates exactly for this modification.
Equation~(\ref{eq:rn-identity}) is exact for every finite $N$. No large-$N$ approximation has been made.

The advantage of this representation depends on choosing a suitable auxiliary density. We want successful propagation to be sufficiently common under $h_a$, while keeping the likelihood weights controlled well enough to determine their exponential contribution. For the optimal Beta proposals
constructed in Section~\ref{sec:beta}, four properties accomplish this.

\medskip
\noindent
\textit{Property 1: Barrier dominance.}

We require
\begin{equation}
H_a(t)\ge\frac{t}{a},
\qquad 0\le t\le a.
\end{equation}
This means that, at every threshold level, the proposal places at least as much probability below that level as the uniform distribution on $(0,a)$.

\medskip
\noindent
\textit{Property 2: Monotone likelihood weights.}

We require $\ell_a(t)$ to be nondecreasing and its grid average to converge to its continuum average:
\begin{equation}
\frac1n\sum_{j=1}^n
\ell_a\left(\frac{aj}{n}\right)
=
\frac1a\int_0^a\ell_a(t)\,dt+o(1).
\label{eq:riemann-cond}
\end{equation}
Monotonicity will allow the propagation inequalities to provide a direct upper bound on the total likelihood weight.

\medskip
\noindent
\textit{Property 3: Equality of spatial and statistical averages.}

For the proposals of interest,
\begin{equation}
\frac1a\int_0^a\ell_a(t)\,dt
=
\int_0^a h_a(t)\ell_a(t)\,dt
=:m_a.
\label{eq:mean-eq}
\end{equation}
This equality is special to the structure of the optimal proposals; it does not hold for an arbitrary change of measure.
The second integral is the average logarithmic weight under $h_a$. By the definition of relative entropy,
\begin{equation}
m_a=-D_{\mathrm{KL}}(h_a\Vert q_a).
\end{equation}
For the optimal proposal, $m_a$ is therefore minus the minimum entropic cost of satisfying the propagation constraint. We also require $m_a\to m_x$ as $N\to\infty$ with $k/N\to x$.

\medskip
\noindent
\textit{Property 4: Exponential concentration.}

Under independent sampling from $h_a$, the average logarithmic weight is expected to approach $m_a$. We require the probability of a fixed downward deviation to decrease exponentially:
\begin{equation}
\mathbb{P}_{h_a}
\left\{
\frac1n\sum_{i=1}^n
\ell_a(Y_i)<m_a-\varepsilon
\right\}
\le e^{-c_\varepsilon n},
\label{eq:chernoff-cond}
\end{equation}
for every $\varepsilon>0$, with some $c_\varepsilon>0$ independent of $N$, provided $N$ is sufficiently large and $k/N\to x\in(0,1)$.
This condition is stronger than the ordinary law of large numbers. It ensures that unusually small likelihood weights are negligible even compared with an event whose probability decreases algebraically with $n$.

For the optimal Beta proposals constructed in the next section, Properties 1 and 3 follow directly from their structure. Property 1 holds because the entropy minimizer belongs to the admissible set $\mathcal{A}_a$. To understand Property 3, note that the optimal proposals consist of free intervals,
where $h_a=q_a$ and hence $\ell_a=0$, and pinned intervals, where $h_a=1/a$. In either case, $h_a(t)\ell_a(t)=\ell_a(t)/a$, which immediately yields Eq.~(\ref{eq:mean-eq}).
Properties 2 and 4 require additional estimates, particularly when the Beta density has endpoint singularities. Their verification is deferred to Appendix~\ref{app:technical}.

\subsection{Parking functions and matching exponential bounds}
\label{subsec:bounds}

We now show how the four properties establish the exponential propagation rate. The argument has two ingredients: a lower bound on the probability
of successful propagation under the proposal distribution, and upper and lower bounds on the associated likelihood weights.

\paragraph{Successful propagation under the proposal.}

Property 1 implies that a random variable drawn from $h_a$ can be coupled to a uniform random variable so that the former is never larger than $a$ times the latter. To see this, let $U$ be uniform on $[0,1]$ and generate
\begin{equation}
Y=H_a^{-1}(U).
\end{equation}
Because $H_a(t)\ge t/a$, the corresponding quantiles satisfy
\begin{equation}
H_a^{-1}(u)\le au,
\qquad 0<u<1.
\end{equation}
Applying this construction to $n$ independent uniform variables gives
\begin{equation}
Y_{(j)}\le aU_{(j)},
\qquad j=1,\ldots,n.
\end{equation}
Therefore, whenever the uniform variables satisfy the parking-function conditions $U_{(j)}\le j/n$, the proposal variables
automatically satisfy $Y_{(j)}\le aj/n$. The parking-function identity~(\ref{eq:parking}) thus provides the bound
\begin{equation}
\mathbb{P}_{h_a}(E_n)
\ge
\frac{(n+1)^{n-1}}{n^n}
\sim\frac{e}{n}.
\label{eq:ph-lower}
\end{equation}
This is the decisive observation. Under the auxiliary distribution, successful propagation is not exponentially rare. Its probability is bounded below by a quantity that decreases only algebraically with $n$.

The remaining task is to determine the exponential contribution of the likelihood weights in Eq.~(\ref{eq:rn-identity}). To sandwich $R_{N,k}$ between matching asymptotic bounds, we must rule out two opposite possibilities: that successful propagation events carry anomalously \emph{large} weights (upper bound) or anomalously \emph{small} weights (lower bound).

\paragraph{Upper bound.}

To understand the upper bound, recall from Property~2 that the log-likelihood ratio $\ell_a(t)=\log[q_a(t)/h_a(t)]$ is nondecreasing. Intuitively, the propagation constraints force the ordered thresholds toward smaller values, where the likelihood ratio is no larger. Thus, satisfying the propagation barrier cannot increase the likelihood weight beyond the value obtained by placing each ordered threshold at its largest permitted value, $Y_{(j)}=aj/n$.

Formally, whenever $E_n$ occurs, we have $Y_{(j)}\le aj/n$ for every $j=1,\ldots,n$. Since $\ell_a$ is nondecreasing,
\begin{equation}
\sum_{i=1}^n\ell_a(Y_i)
=
\sum_{j=1}^n\ell_a(Y_{(j)})
\le
\sum_{j=1}^n
\ell_a\left(\frac{aj}{n}\right)
=nm_a+o(n),
\end{equation}
where the last step follows from Properties~2 and~3.

Thus, on every successful propagation event, the likelihood weight is bounded deterministically from above by $e^{nm_a+o(n)}$. Substituting this bound into Eq.~(\ref{eq:rn-identity}) gives
\begin{equation}
R_{N,k}
\le
e^{nm_a+o(n)}
\mathbb{P}_{h_a}(E_n)
\le e^{nm_a+o(n)}.
\end{equation}
Taking logarithms, dividing by $N$, and using $n/N\to x$ and $m_a\to m_x$, we obtain
\begin{equation}
\limsup_{N\to\infty}
\frac1N\log R_{N,k}
\le xm_x.
\label{eq:upper-lim}
\end{equation}

\paragraph{Lower bound.}

For the lower bound, we restrict attention to realizations whose likelihood weights are not unusually small. Fix a tolerance $\varepsilon>0$ and define $B_{n,\varepsilon}$ as the set of threshold configurations $(Y_1,\ldots,Y_n)$ for which the sample-averaged log-likelihood ratio does not fall more than $\varepsilon$ below its expectation $m_a$:
\begin{equation}
B_{n,\varepsilon}
=
\left\{
\frac1n\sum_{i=1}^n
\ell_a(Y_i)
\ge m_a-\varepsilon
\right\}.
\end{equation}
Thus, $B_{n,\varepsilon}$ contains all realizations whose likelihood weights are not anomalously small, whereas its complement $B_{n,\varepsilon}^c$ consists of configurations with an unusually large downward deviation of the average log-likelihood ratio.

On the intersection $E_n\cap B_{n,\varepsilon}$, propagation succeeds ($\mathbf{1}_{E_n}=1$) and the likelihood weight is guaranteed to be at least $e^{n(m_a-\varepsilon)}$. Because the integrand in Eq.~(\ref{eq:rn-identity}) is everywhere non-negative, restricting the expectation to $E_n\cap B_{n,\varepsilon}$ yields
\begin{equation}
R_{N,k}
\ge
\mathbb{E}_{h_a}\!\left[
\mathbf{1}_{E_n\cap B_{n,\varepsilon}}
\exp\!\left(\sum_{i=1}^n\ell_a(Y_i)\right)
\right]
\ge
e^{n(m_a-\varepsilon)}
\mathbb{P}_{h_a}
(E_n\cap B_{n,\varepsilon}).
\label{eq:lower-step1}
\end{equation}
To estimate the remaining probability, we use the elementary inequality
\begin{equation}
\mathbb{P}_{h_a}
(E_n\cap B_{n,\varepsilon})
\ge
\mathbb{P}_{h_a}(E_n)
-
\mathbb{P}_{h_a}
(B_{n,\varepsilon}^{c}).
\end{equation}
Importantly, this inequality does not require $E_n$ and $B_{n,\varepsilon}$ to be independent.
The first probability is bounded below by the parking-function result, while the second is exponentially small by Property~4. Hence
\begin{equation}
\mathbb{P}_{h_a}
(E_n\cap B_{n,\varepsilon})
\ge
\frac{(n+1)^{n-1}}{n^n}
-e^{-c_\varepsilon n}.
\end{equation}

The crucial point is that the probability of successful propagation under the auxiliary measure decreases at most algebraically, whereas the probability
of anomalously small likelihood weights decreases exponentially. Therefore, the latter cannot remove a significant fraction of the former on the exponential scale.

Indeed, for sufficiently large $n$, the algebraic term dominates the exponentially small correction, implying
\begin{equation}
\lim_{N\to\infty}
\frac1N\log
\mathbb{P}_{h_a}
(E_n\cap B_{n,\varepsilon})
=0.
\end{equation}
Using this result in Eq.~(\ref{eq:lower-step1}) gives
\begin{equation}
\liminf_{N\to\infty}
\frac1N\log R_{N,k}
\ge x(m_x-\varepsilon).
\end{equation}
Since $\varepsilon>0$ can be chosen arbitrarily small, we conclude that
\begin{equation}
\liminf_{N\to\infty}
\frac1N\log R_{N,k}
\ge xm_x.
\label{eq:lower-lim}
\end{equation}

The upper and lower bounds coincide. We have therefore established
\begin{equation}
\lim_{N\to\infty}
\frac1N\log R_{N,k_N}
=xm_x.
\label{eq:tilt-rate}
\end{equation}

For the optimal proposal $h_x=g_x^*$, we have $m_x=-\mathcal{I}_F(x)$. Combining this result with the factorization $Q_{N,k}=F(a)^nR_{N,k}$ finally gives
\begin{equation}
\lim_{N\to\infty}
\frac1N\log Q_{N,k_N}
=
x\log F(x)-x\mathcal{I}_F(x).
\end{equation}
This establishes the continuum prediction (\ref{eq:psi-candidate}) whenever the four properties of the proposal density hold.

The methodological significance extends beyond threshold models. The entropy minimizer identifies the most probable realization of a rare event, while the change of measure makes that realization accessible at finite sample size. The parking-function coupling ensures that the constrained event is not exponentially rare under the new distribution, and the likelihood bounds recover its original exponential cost.

In the next section, we construct the optimal proposal densities explicitly for Beta threshold distributions and verify their barrier and average properties
(Properties 1 and 3). The remaining technical requirements (Properties 2 and 4) are established in Appendix~\ref{app:technical}.


\section{Beta densities and four optimal profiles}
\label{sec:beta}

We now apply the constrained-entropy method of Section~\ref{sec:entropy} to the family of Beta threshold distributions. Our objective is to determine the most probable threshold profile compatible with a macroscopic cascade of size $x$, and from this profile calculate the exponential propagation rate $\psi_Q(x)$.

The main advantage of the Beta family is that its density can be decreasing, increasing, bell-shaped, or U-shaped. These four possibilities lead to different ways of satisfying the propagation constraint and therefore to different optimal profiles.

Throughout this section, the optimal profile $g_x^*$ minimizes the relative entropy
$D_{\mathrm{KL}}(g\Vert q_x)$, Eq.~(\ref{eq:KL_qg}), subject to
\begin{equation}
G(t)=\int_0^t g(s)\,ds\ge\frac{t}{x},
\qquad 0\le t\le x.
\end{equation}
Once this minimizer is known, the propagation rate follows from Eq.~(\ref{eq:psi-candidate}):
\begin{equation}
\psi_Q(x)
=
x\log F(x)
-xD_{\mathrm{KL}}(g_x^*\Vert q_x).
\label{eq:beta-rate-general}
\end{equation}

The calculations below identify the optimal profiles and evaluate their entropic costs. Their optimality is established in Appendix~\ref{app:beta-optimality}, while the existence and uniqueness of the matching points are discussed in Appendix~\ref{app:beta-geometry}.

\subsection{The role of the shape of the density}

The Beta density and its cumulative distribution are
\begin{equation}
\label{eq:beta}
f(t)=
\frac{t^{\alpha-1}(1-t)^{\beta-1}}
{\mathrm{B}(\alpha,\beta)},
\qquad
F(t)=\int_0^t f(s)\,ds,
\qquad \alpha,\beta>0.
\end{equation}
The qualitative behavior of $f$ is governed by its logarithmic derivative,
\begin{equation}
\frac{f'(t)}{f(t)}
=
\frac{\alpha-1}{t}
-
\frac{\beta-1}{1-t}.
\label{eq:beta-log-derivative}
\end{equation}
Depending on the values of $\alpha$ and $\beta$, the density may favor low thresholds, high thresholds, intermediate thresholds, or both extremes.

This shape directly determines whether the conditioned threshold distribution $q_x(t)=f(t)/F(x)$ can sustain sequential recruitment without further modification. If $q_x$ already places sufficient probability mass at small thresholds so that its cumulative distribution satisfies $F(t)/F(x)\ge t/x$ throughout $[0,x]$, no rearrangement is required and $g_x^*=q_x$. Conversely, when this condition fails, sustaining a cascade of size $x$ requires an atypical redistribution of thresholds toward smaller values, with an associated entropic cost.

Minimizing the relative entropy $D_{\mathrm{KL}}(g\Vert q_x)$ subject to normalization and the propagation constraint $G(t)\ge t/x$ yields an optimal profile $g_x^*$ with a simple piecewise structure. For the Beta family, the interval $[0,x]$ contains at most one free interval and one pinned interval:

\begin{enumerate}[label=(\roman*)]
\item On a \emph{free interval}, the cumulative constraint is inactive ($G(t)>t/x$ in the interior). The stationarity condition for relative-entropy minimization then requires the ratio $g_x^*(t)/q_x(t)$ to be constant. Consequently, the optimal density retains the shape of the original distribution: $g_x^*(t)\propto f(t)$.

\item On a \emph{pinned interval}, the cumulative distribution saturates the propagation boundary, $G(t)=t/x$, which forces the density to be uniform: $g_x^*(t)=1/x$.
\end{enumerate}

For the Beta profiles considered here, the proportionality constants and contact points are determined by normalization and matching conditions at the boundaries between free and pinned intervals. The resulting optimal densities are continuous at these boundaries.

This free/pinned decomposition provides a unified framework for all four parameter regimes. Rather than solving a separate variational problem in each case, classifying the large-deviation behavior reduces to identifying which intervals are pinned, which remain free, and how their contact points move as $x$ varies.

For later use, define
\begin{align}
\label{eq:J}
J(c)
&=\int_0^c\log f(t)\,dt \nonumber \\
&=(\alpha-1)(c\log c-c) -(\beta-1)
\big[(1-c)\log(1-c)+c\big]
-c\log\mathrm{B}(\alpha,\beta).
\end{align}
At $c=0$, the expression $c\log c$ is understood by continuity. The function $J(c)$ allows us to write the propagation rates in closed form once the optimal profile has been identified.

\subsection{Decreasing densities: no rearrangement is needed}

Consider first
\begin{equation}
\alpha\le1,\qquad \beta\ge1,
\end{equation}
for which the Beta density is nonincreasing. In this regime, low thresholds are at least as probable as higher ones. We therefore expect the conditioned
distribution $q_x$ to contain enough low-threshold individuals to satisfy the propagation constraint without any additional modification.

This expectation can be checked directly. Since $f$ is nonincreasing, its average over $[0,t]$ is at least its average over $[0,x]$ whenever
$t\le x$:
\begin{equation}
\frac{F(t)}{t}
\ge
\frac{F(x)}{x}.
\end{equation}
Equivalently,
\begin{equation}
\frac{F(t)}{F(x)}
\ge\frac{t}{x}.
\end{equation}
But $F(t)/F(x)$ is precisely the cumulative distribution of $q_x$. Thus, the conditioned density already satisfies the propagation barrier.

Since relative entropy is nonnegative and vanishes when the two distributions coincide, there is no advantage in changing the distribution. The optimal profile is simply
\begin{equation}
g_x^*(t)=q_x(t)=\frac{f(t)}{F(x)}.
\end{equation}
Its entropic cost is zero, and Eq.~(\ref{eq:beta-rate-general}) immediately gives
\begin{equation}
\psi_Q(x)=x\log F(x).
\label{eq:rate-I}
\end{equation}
This is the simplest possible outcome of the variational problem. The leading exponential cost comes entirely from requiring all recruited thresholds to lie
below $x$. The additional sequential propagation constraints introduce no further exponential penalty.

\subsection{Increasing densities: the entire profile is pinned}

Now consider
\begin{equation}
\alpha\ge1,\qquad \beta\le1,
\end{equation}
for which $f$ is nondecreasing. Here the situation is reversed. Higher thresholds are favored, so the conditioned distribution places relatively little probability near the origin. Except for the uniform case, it cannot sustain sequential recruitment without a redistribution of thresholds toward smaller values.

The optimal rearrangement is particularly simple: the cumulative distribution follows the propagation boundary over the entire interval,
\begin{equation}
G_x^*(t)=\frac{t}{x}.
\end{equation}
Differentiating gives
\begin{equation}
g_x^*(t)=\frac1x,
\qquad 0<t<x.
\end{equation}
Thus, the entire interval is pinned. The optimal profile is uniform, even though the original density is generally increasing.

The emergence of a uniform optimal profile reflects a balance between minimizing entropic cost and satisfying the recruitment barrier at every stage. Because the original density is nondecreasing, shifting probability mass toward smaller thresholds incurs an entropic penalty. The propagation constraint requires a minimum cumulative supply of low-threshold individuals, represented by the boundary $G(t)=t/x$. Any admissible profile lying strictly above this boundary contains an additional excess of low-threshold individuals.
For a nondecreasing original density, such an excess cannot reduce the relative-entropy cost. The least costly admissible profile therefore saturates the propagation constraint throughout the interval. This conclusion follows rigorously from the relative-entropy optimality argument given in Appendix~\ref{app:beta-optimality}.

To calculate the rate, substitute $g_x^*=1/x$ into the relative entropy:
\begin{equation}
D_{\mathrm{KL}}(g_x^*\Vert q_x)
=
\frac1x\int_0^x
\log\frac{F(x)}{xf(t)}\,dt.
\end{equation}
Using the definition of $J(x)$, we obtain
\begin{equation}
D_{\mathrm{KL}}(g_x^*\Vert q_x)
=
\log F(x)-\log x-\frac{J(x)}{x}.
\end{equation}
Equation~(\ref{eq:beta-rate-general}) then gives
\begin{equation}
\psi_Q(x)=x\log x+J(x).
\label{eq:rate-II}
\end{equation}

Unlike the decreasing-density case, successful propagation generally requires an extensive reorganization of the threshold distribution. The resulting entropy cost changes the leading exponential propagation rate.

Two limiting cases provide useful connections with earlier exact results. First, for $\alpha=1$ and $0<\beta<1$, Zeraoulia~\cite{zeraoulia_comment} obtained the limiting distribution of microscopic cascades, in which the final cascade size $k$ remains fixed as $N\to\infty$. In particular, the propagation
factor satisfies
\begin{equation}
Q_{N,k}\sim
\frac{\beta^{k-1}k^{k-2}}{N^{k-1}},
\qquad k\ge2.
\end{equation}
Our result instead describes macroscopic cascades, for which $k/N\to x\in(0,1)$. Nevertheless, the two asymptotic descriptions are consistent at the level of their leading exponential dependence. Indeed, for $\alpha=1$ and $x\to0$,
\begin{equation}
\psi_Q(x)
=x\log(\beta x)+O(x^2).
\end{equation}
Formally setting $x=k/N$ gives the factor $(\beta k/N)^k$, which has the same leading exponential dependence as the microscopic expression. The remaining algebraic factors cannot be recovered from the macroscopic  rate alone, since the limits $k=O(1)$ and $k/N\to x>0$ describe different asymptotic regimes.

A second benchmark is the uniform distribution, $\alpha=\beta=1$, discussed in Section~\ref{subsec:parking}. In this case, the parking-function identity gives the exact finite-population expression
\begin{equation}
Q_{N,k}=\frac{k^{k-2}}{N^{k-1}},
\end{equation}
and hence $\psi_Q(x)=x\log x$, precisely as predicted by Eq.~(\ref{eq:rate-II}). The uniform case thus provides an exact connection between the finite-population combinatorics and the macroscopic large-deviation rate.

\subsection{Bell-shaped densities: a pinned region followed by a free region}

Consider next
\begin{equation}
\alpha>1,\qquad \beta>1.
\end{equation}
The Beta density is now bell-shaped, increasing up to its mode
\begin{equation}
m=\frac{\alpha-1}{\alpha+\beta-2}
\end{equation}
and decreasing thereafter. This change in monotonicity suggests that the optimal profile may combine the two mechanisms encountered above.

For cascade sizes $x\le m$, the density is increasing throughout the relevant interval $[0,x]$. The argument of the preceding subsection therefore applies: the entire optimal profile is pinned,
\begin{equation}
g_x^*(t)=\frac1x,
\qquad x\le m,
\end{equation}
and
\begin{equation}
\psi_Q(x)=x\log x+J(x).
\end{equation}

The more interesting situation occurs when $x>m$. The original density now increases over part of the interval and decreases over the remainder. We therefore expect the propagation constraint to be active at small thresholds, where the original density is increasing, but to become inactive farther along the interval.

This leads to a profile with a pinned region $0<t\le c_x$ and a free region $c_x<t<x$:
\begin{equation}
\label{eq:g-bell}
g_x^*(t)=
\begin{cases}
\dfrac1x,
&0<t\le c_x,\\[2mm]
\dfrac{f(t)}{xf(c_x)},
&c_x<t<x.
\end{cases}
\end{equation}
The matching point $c_x$ must satisfy $0<c_x<m$. Indeed, for $G_x^*(t)$ to rise above the propagation barrier $t/x$ immediately after leaving it at $t=c_x$, the free-region density must exceed $1/x$ for $t$ slightly larger than $c_x$. This requires $f(t)>f(c_x)$ just beyond $c_x$, which is possible only on the increasing side of the mode ($c_x<m$).

The proportionality constant in the free region is fixed by requiring continuity of the optimal density at $c_x$: $g_x^*(c_x)=1/x$. This explains the factor
$1/[xf(c_x)]$ in the second branch of Eq.~(\ref{eq:g-bell}). The remaining unknown, the location $c_x$ itself, is determined by requiring the complete profile to be normalized.

The normalization condition reads
\begin{equation}
\int_0^{c_x}\frac{dt}{x}
+
\int_{c_x}^x
\frac{f(t)}{xf(c_x)}\,dt
=1,
\end{equation}
which gives
\begin{equation}
F(x)-F(c_x)=(x-c_x)f(c_x).
\label{eq:c-bell}
\end{equation}
This equation has a simple geometrical meaning. The area under the original density between $c_x$ and $x$ must equal the area of a rectangle with width $x-c_x$ and height $f(c_x)$. The matching condition ensures that the free part contains exactly the probability mass needed to complete the
normalized distribution.

For every $x>m$, there is a unique solution $c_x\in(0,m)$. The corresponding cumulative distribution lies on the propagation boundary for $t\le c_x$, rises above it for $c_x<t<x$, and meets it again at $t=x$. The proof of these statements is given in Appendix~\ref{app:beta-geometry}.

We can now evaluate the propagation rate. On the pinned interval, $g_x^*=1/x$, whereas on the free interval $g_x^*=f(t)/[xf(c_x)]$. Using the variational
expression for $\psi_Q$, the two contributions are
\begin{equation}
\psi_Q(x)
=
\int_0^{c_x}
\log[xf(t)]\,dt
+
(x-c_x)\log[xf(c_x)].
\end{equation}
The first term is the contribution of the pinned region. The second comes from the free region, where the ratio between the original and optimal densities is constant. Collecting terms gives
\begin{equation}
\psi_Q(x)
=
x\log x+J(c_x)
+(x-c_x)\log f(c_x).
\label{eq:rate-III}
\end{equation}
For convenience, we may set $c_x=x$ when $x\le m$. With this convention, Eq.~(\ref{eq:rate-III}) also describes the fully pinned regime.

\subsection{U-shaped densities: a free region followed by a pinned region}

Finally, consider
\begin{equation}
\alpha<1,\qquad \beta<1.
\end{equation}
The Beta density is U-shaped. It decreases from the origin to its minimum at
\begin{equation}
m=\frac{1-\alpha}{2-\alpha-\beta}
\end{equation}
and then increases toward the upper endpoint. The abundance of very small thresholds initially favors propagation. However, the increasing part of the density eventually creates a different situation: the conditioned distribution may no longer contain enough low-threshold individuals to satisfy the propagation barrier throughout $[0,x]$. We therefore expect two regimes, separated by a characteristic value $d$.

To determine the value $d$, consider the cumulative distribution of the conditioned density,
\begin{equation}
G(t)=\frac{F(t)}{F(x)}.
\end{equation}
The propagation constraint is equivalent to
\begin{equation}
\frac{F(t)}{t}
\ge\frac{F(x)}{x},
\qquad 0<t\le x.
\end{equation}
Thus, the relevant quantity is the average density $F(t)/t$. Its derivative is
\begin{equation}
\frac{d}{dt}\frac{F(t)}{t}
=
\frac{tf(t)-F(t)}{t^2}.
\end{equation}
For a U-shaped Beta density, this average first decreases and then increases. Its minimum occurs at a unique point $d>m$ satisfying
\begin{equation}
F(d)=df(d).
\label{eq:d}
\end{equation}
Geometrically, the density $f(d)$ equals its average value over the interval $[0,d]$.
The existence and uniqueness of $d$ are established in Appendix~\ref{app:beta-geometry}.

For $x\le d$, the average density $F(t)/t$ is nonincreasing throughout $[0,x]$. Consequently, the conditioned distribution already satisfies the propagation barrier:
\begin{equation}
\frac{F(t)}{F(x)}
\ge\frac{t}{x}.
\end{equation}
No rearrangement is needed, and the optimal profile is
\begin{equation}
g_x^*(t)=q_x(t),
\qquad x\le d.
\end{equation}
The corresponding rate is
\begin{equation}
\psi_Q(x)=x\log F(x).
\end{equation}

For $x>d$, however, the conditioned density is no longer admissible. The optimal profile must be modified. In contrast with the bell-shaped case, the
free region now comes first, followed by a pinned region:
\begin{equation}
\label{eq:g-U}
g_x^*(t)=
\begin{cases}
\dfrac{f(t)}{xf(d)},
&0<t\le d,\\[2mm]
\dfrac1x,
&d<t<x.
\end{cases}
\end{equation}
The free part retains the shape of the original density, including its strong concentration near the origin. The pinned part follows the propagation boundary over the remaining interval. The matching condition $F(d)=df(d)$ guarantees normalization:
\begin{equation}
\frac{F(d)}{xf(d)}
+\frac{x-d}{x}
=1.
\end{equation}
It also ensures that the cumulative distribution reaches the propagation boundary at $t=d$. Before that point, the optimal cumulative distribution lies above the boundary; afterward, it coincides with it.

An interesting feature of this regime is that the matching point $d$ does not depend on the final cascade size $x$. Once $x>d$, increasing the cascade size extends the pinned interval but leaves the free interval unchanged.

The rate is obtained by adding the contributions from the free and pinned regions:
\begin{equation}
\psi_Q(x)
=
d\log[xf(d)]
+
\int_d^x\log[xf(t)]\,dt.
\end{equation}
Using the definition of $J$, we obtain
\begin{equation}
\psi_Q(x)
=
x\log x+d\log f(d)
+J(x)-J(d),
\qquad x>d.
\label{eq:rate-IV}
\end{equation}

The U-shaped family therefore undergoes a different change in the optimal profile: for $x\le d$ the original conditioned distribution is already admissible, whereas for $x>d$ a pinned region appears at the upper end of the interval.

\begin{figure}[t]
\centering
\includegraphics[width=\textwidth]{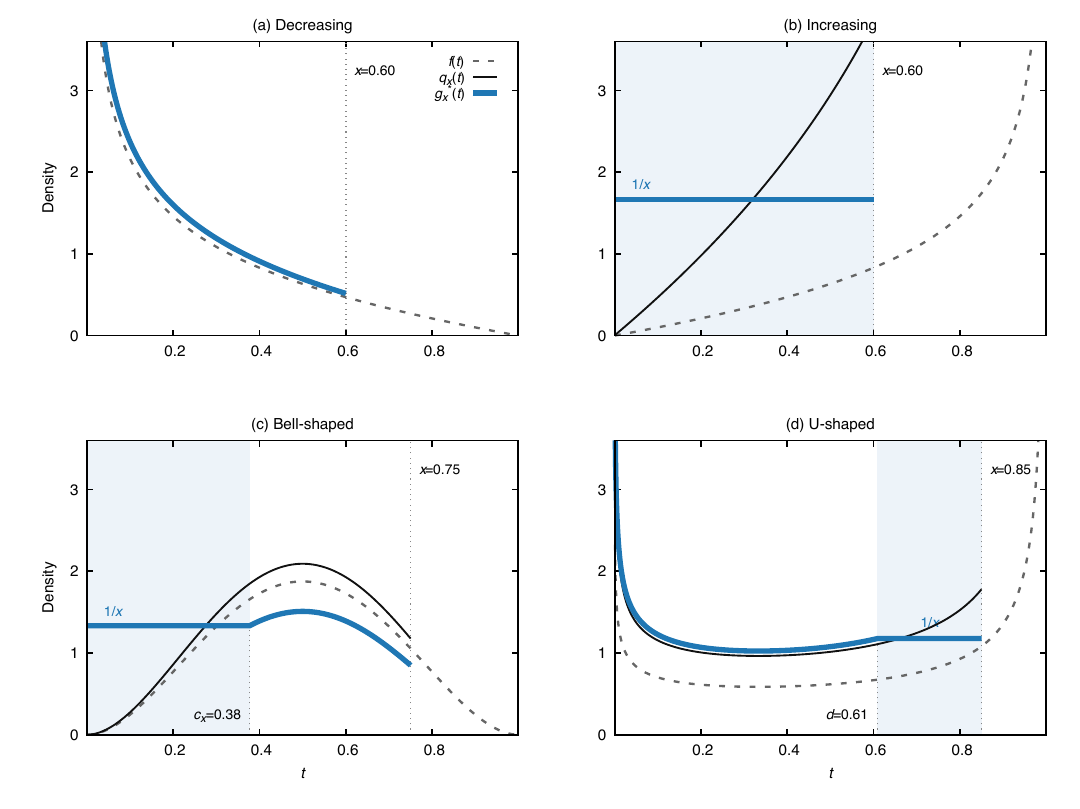}
\caption{
Four regimes of the entropy-minimizing threshold density $g_x^*(t)$ for Beta-distributed thresholds. The dashed gray curves represent the original density $f(t)$, the thin black curves the conditioned density $q_x(t)=f(t)/F(x)$, and the thick blue curves the optimal profile $g_x^*(t)$. (a) Decreasing density: $\alpha=0.6$, $\beta=2$, $x=0.60$; the entire interval is free. (b) Increasing density: $\alpha=2$, $\beta=0.6$, $x=0.60$;
the entire interval is pinned. (c) Bell-shaped density: $\alpha=\beta=3$, $x=0.75$; a pinned interval is followed by a free interval, with $c_x\approx0.38$. (d) U-shaped density: $\alpha=0.7$, $\beta=0.4$, $x=0.85$; a free interval is followed by a pinned interval, with $d\approx0.61$. Light shading indicates pinned intervals, where $g_x^*(t)=1/x$. The conditioned and optimal densities are defined only for $0<t<x$.
}
\label{fig:optimal_profiles}
\end{figure}

Figure~\ref{fig:optimal_profiles} illustrates the four possible geometries of the entropy-minimizing threshold distributions. The comparison with the conditioned density $q_x$ shows where the propagation constraint requires a redistribution of thresholds and where the original density profile can retain its shape. Importantly, on a free interval the optimal density is proportional to $q_x$, but need not coincide with it. This is because the proportionality constant is determined by the normalization of the entire optimal distribution, including any pinned intervals. Only when the entire interval is free do we have $g_x^*=q_x$, as illustrated in panel (a).

These explicit solutions also complete the finite-sample argument of Section~\ref{sec:entropy}: they satisfy the propagation barrier (Property 1), and their
free/pinned structure ensures the equality of likelihood averages (Property 3). The remaining technical conditions (Properties 2 and 4) are verified in Appendix~\ref{app:technical}. Consequently, the rates derived above are the exact leading exponential rates of the finite-population propagation probabilities.


\subsection{Final cascade-size rate}
\label{sec:final-rate}

The preceding analysis determines the leading exponential rate of the propagation factor $Q_{N,k}$. Using Eq.~(\ref{eq:P}) and Stirling's formula, we obtain the corresponding large-deviation rate of the final cascade-size distribution,
\begin{equation}
\label{eq:IP}
I_P(x)= -
\lim_{N\to\infty}
\frac1N\log P_N(k)
=
x\log x+(1-x)\log(1-x)
-\psi_Q(x)
-(1-x)\log[1-F(x)],
\end{equation}
where $k/N\to x\in(0,1)$. Thus, the explicit propagation rates derived in the previous section also determine the leading exponential probabilities of macroscopic final cascade sizes.

The uniform case $\alpha=\beta=1$, discussed in Section~\ref{subsec:parking}, provides an exact benchmark. The parking-function
identity gives $\psi_Q(x)=x\log x$ and $I_P(x)=0$, although the probability of any particular macroscopic cascade size decays as $N^{-3/2}$. This illustrates that a vanishing large-deviation rate does not imply a nonzero limiting probability.


\section{Numerical verification and optimal empirical profiles}
\label{sec:numerics}

We test the analytical large-deviation rates and optimal empirical threshold profiles derived in Section~\ref{sec:beta} for two representative distributions: $\mathrm{Beta}(2,2)$ (bell-shaped) and $\mathrm{Beta}(1/2,1/2)$ (U-shaped). We use importance sampling to estimate the propagation rates and sequential Monte Carlo to examine the threshold distributions conditioned on successful macroscopic cascades.

For the importance-sampling estimates, we fix $k=\lfloor xN\rfloor$, $n=k-1$, and $a=n/N$, and generate thresholds from the optimal proposal $h_a=g_a^*$ derived in Section~\ref{sec:beta}. Using the propagation event $E_n$ defined in Eq.~(\ref{eq:E}), the conditional propagation probability is estimated by
\begin{equation}
\widehat R_{N,k}
=\frac{1}{M}\sum_{m=1}^{M}
\mathbf{1}_{E_n}^{(m)}
\prod_{i=1}^{n}\frac{q_a(Y_i^{(m)})}
{h_a(Y_i^{(m)})},
\label{eq:IS_estimator}
\end{equation}
where $M$ is the number of independent samples. The finite-population rate estimate follows from the exact factorization~(\ref{eq:factor}):
\begin{equation}
\widehat\psi_{Q,N}(x)
=\frac{n}{N}\log F(a)
+\frac{1}{N}\log\widehat R_{N,k}.
\label{eq:numerical_rate}
\end{equation}
We use $M=20000$ samples for each $x$ and $N=100$, $400$, and $1600$, accumulating likelihood ratios logarithmically to avoid numerical underflow.

To sample the empirical threshold distributions conditioned on propagation, we employ a separate sequential Monte Carlo algorithm. The ordered
thresholds are generated successively under $q_a$, enforcing the propagation constraints at each step. Given $Y_{(j-1)}=y$, the probability that the next order statistic satisfies its constraint is
\begin{equation}
p_j(y)=1-
\left[
\frac{F(a)-F(j/N)}
     {F(a)-F(y)}
\right]^{n-j+1},
\label{eq:SMC_probability}
\end{equation}
with $y=0$ for $j=1$. Particle weights are updated by $p_j(y)$, with systematic resampling when the effective sample size falls below a prescribed threshold. Independent runs are combined using their estimated propagation probabilities as weights. Since this method does not use the analytical optimizer $g_a^*$, it provides an independent test of the predicted conditional profiles.

\begin{figure}[t]
\centering
\includegraphics[width=\textwidth]{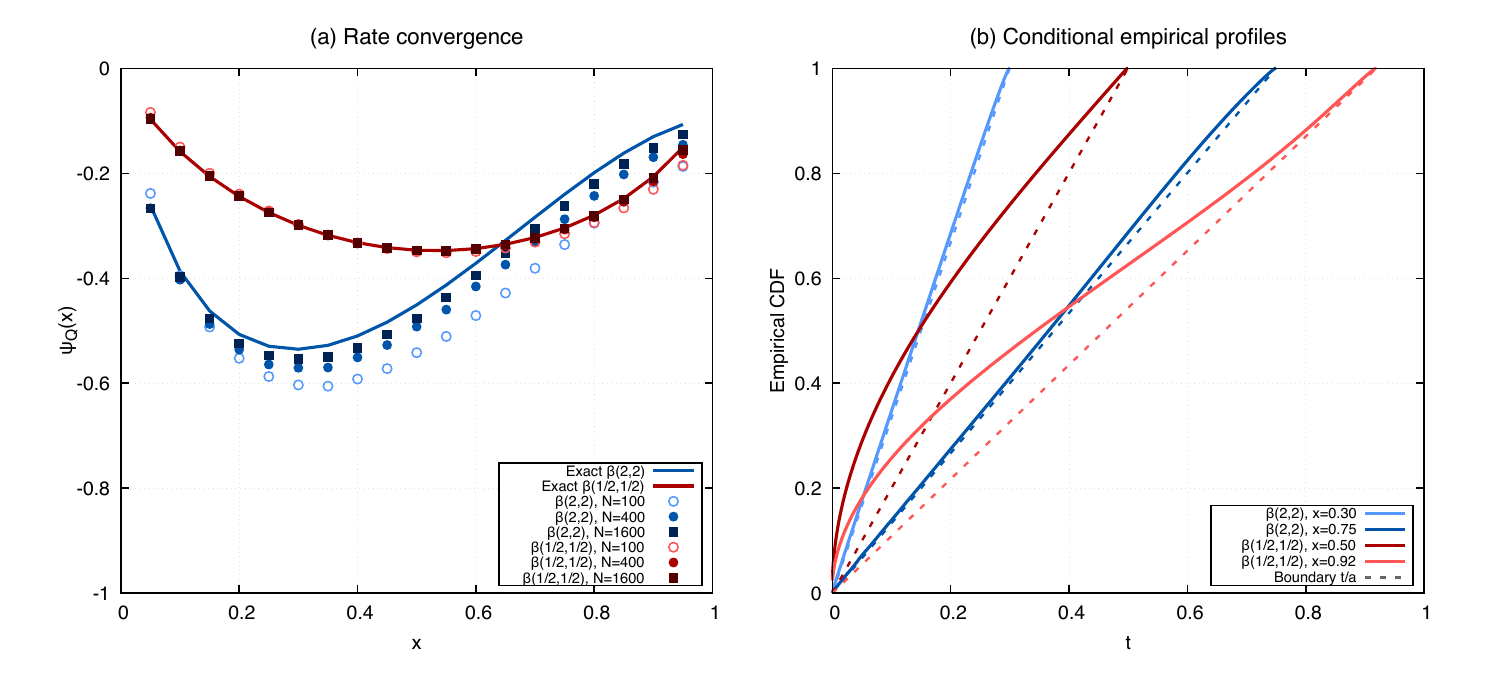}
\caption{
Numerical verification of the propagation rates and conditional empirical threshold profiles. (a) Finite-population estimates of $N^{-1}\log Q_{N,k}$ for $\mathrm{Beta}(2,2)$ (blue) and $\mathrm{Beta}(1/2,1/2)$ (red), with $k=\lfloor xN\rfloor$. Open circles, filled circles, and squares correspond to $N=100$, $400$, and $1600$, respectively. Solid curves are the analytical asymptotic rates. Each estimate uses $20000$ importance-sampling samples. (b) Conditional empirical threshold CDFs for $N=1600$, obtained by sequential Monte Carlo. The curves correspond to $\mathrm{Beta}(2,2)$
at $x=0.30$ and $0.75$, and $\mathrm{Beta}(1/2,1/2)$ at $x=0.50$ and $0.92$. Dashed lines indicate the propagation boundary $G(t)=t/a$, with $a=(k-1)/N$.
}
\label{fig:numerical}
\end{figure}

Figure~\ref{fig:numerical}(a) shows that the finite-population rate estimates approach the analytical predictions as $N$ increases, with more pronounced finite-size deviations for the bell-shaped distribution. Figure~\ref{fig:numerical}(b) confirms the four predicted optimal-profile geometries: fully pinned and pinned--free for $\mathrm{Beta}(2,2)$, and fully free and free--pinned for $\mathrm{Beta}(1/2,1/2)$. The agreement provides complementary evidence
for both the exponential propagation rates and the threshold configurations responsible for rare macroscopic cascades.


\section{Critical scaling near $\alpha=1$}
\label{sec:critical-scaling}

We begin by recalling the deterministic dynamics in the infinite-population limit. Starting from a single instigator, whose initial fraction vanishes as $N\to\infty$, the fraction of active individuals evolves according to $x_{t+1}=F(x_t)$. Its fixed points satisfy $x^*=F(x^*)$, and $x^*=0$ is always a solution. The stability of this solution is determined by $F'(0^+)=f(0^+)$: it is stable when $f(0^+)<1$ and unstable when $f(0^+)>1$.

For the $\mathrm{Beta}(\alpha,\beta)$ family, $f(0^+)=0$ when $\alpha>1$ and diverges when $\alpha<1$, whereas $f(0^+)=\beta$ at $\alpha=1$. Thus, for $0<\beta\le1$,
$\alpha_c=1$ separates the regimes in which an infinitesimal initial perturbation dies out or can trigger macroscopic propagation. At $\alpha=1$, the fixed point $x^*=0$ is marginal for $\beta=1$ but remains linearly stable for $0<\beta<1$.

We now examine how this transition is reflected in the finite-size scaling of the mean final cascade fraction,
\begin{equation}
\rho_\infty(N)=\frac{1}{N}
\sum_{k=1}^{N}kP_N(k).
\label{eq:mean-cascade}
\end{equation}
This problem was previously investigated in Ref.~\cite{fontanari2026} for power-law thresholds ($\beta=1$) and Beta distributions with $0<\beta<1$. However, the large-deviation
analysis used there relied on the approximation $\psi_Q(x)=x\log F(x)$, which is not generally valid when the optimal empirical threshold profile contains a pinned interval.

The corrected propagation rate obtained here motivates a reexamination of these scaling results. In particular, macroscopic cascades, with $k/N\to x>0$, are governed by the large-deviation rate $I_P(x)$, whereas microscopic cascades, with $k$ remaining finite as $N\to\infty$, require a separate analysis of the early recruitment process. Both regimes can contribute to Eq.~(\ref{eq:mean-cascade}), and their relative importance depends on the threshold distribution.

\subsection{Power-law thresholds: $\beta=1$}

For $\beta=1$, the cumulative threshold distribution is $F(x)=x^\alpha$. The transition occurs at $\alpha_c=1$, where the threshold distribution is
uniform.

We first consider the macroscopic large-deviation
rate near the transition. Writing $\alpha=1+\epsilon$, with $|\epsilon|\ll1$, the corrected rate depends on the sign of $\epsilon$.

For $\epsilon<0$, the threshold density is decreasing and the optimal profile is fully free. Consequently, $\psi_Q(x)=x\log F(x)$, and the final cascade-size rate reduces to the binary relative entropy
\begin{equation}
I_P(x)=x\log\frac{x}{F(x)}
 +(1-x)\log\frac{1-x}{1-F(x)}.
\end{equation}
Expanding at fixed $0<x<1$ gives
\begin{equation}
I_P(x)=
\frac{x(\log x)^2}{2(1-x)}\epsilon^2
+O(|\epsilon|^3),
\qquad \epsilon<0.
\label{eq:power-left}
\end{equation}

For $\epsilon>0$, the density is increasing and the optimal profile is fully pinned. The propagation rate is instead
\begin{equation}
\psi_Q(x)=x\log x+
\int_0^x\log(\alpha t^{\alpha-1})\,dt,
\end{equation}
which can be evaluated explicitly as
\begin{equation}
\psi_Q(x)=x\log x+x\log\alpha
+(\alpha-1)(x\log x-x).
\end{equation}
Substitution into the exact expression for $I_P(x)$ and expansion around $\alpha=1$ yield
\begin{equation}
I_P(x)=
\frac{x}{2}
\left[
1+\frac{(\log x)^2}{1-x}
\right]\epsilon^2
+O(\epsilon^3),
\qquad \epsilon>0.
\label{eq:power-right}
\end{equation}

Equations~(\ref{eq:power-left}) and (\ref{eq:power-right}) show that the macroscopic rate is quadratic in $\alpha-1$ on both sides of the transition, but with different coefficients. In particular, the pinned regime contributes an additional term $x\epsilon^2/2$ that is absent from the approximation used in
Ref.~\cite{fontanari2026}.

For fixed macroscopic $x$, the exponential suppression therefore becomes appreciable when $N(\alpha-1)^2=O(1)$, identifying the scale
\begin{equation}
|\alpha-1|\sim N^{-1/2}.
\label{eq:power-window}
\end{equation}
This conclusion concerns the macroscopic large-deviation envelope; the scaling of the mean cascade size also depends on its subexponential prefactors.

At the critical point $\alpha=\beta=1$, the propagation factor is known exactly,
\begin{equation}
Q_{N,k}=\frac{k^{k-2}}{N^{k-1}}.
\end{equation}
The resulting final cascade-size distribution satisfies
\begin{equation}
kP_N(k)=
\binom{N}{k}
\left(\frac{k}{N}\right)^k
\left(1-\frac{k}{N}\right)^{N-k}.
\end{equation}
Consequently,
\begin{equation}
\rho_\infty(N)
\sim\sqrt{\frac{\pi}{2N}}.
\label{eq:power-critical}
\end{equation}
This exact result remains unchanged by the correction to the macroscopic propagation rate. The exponent $1/2$ is therefore established independently of the large-deviation approximation. The corrected theory preserves the macroscopic scale $N^{-1/2}$ while revealing an asymmetry between the two sides of the transition.

\subsection{Beta thresholds with $0<\beta<1$}

The situation is qualitatively different when $0<\beta<1$. In this case the transition at $\alpha=1$ is continuous, and the nonzero deterministic fixed point approaches the origin through an essential singularity.

Indeed, for small $x$, the Beta cumulative distribution has the expansion
\begin{equation}
F(x)=\frac{x^\alpha}{\alpha B(\alpha,\beta)}
[1+O(x)].
\label{eq:beta-small-x}
\end{equation}
Writing $\alpha=1+\epsilon$ and solving $F(\rho_\infty)=\rho_\infty$ gives, for
$\epsilon\to0^-$,
\begin{equation}
\log\rho_\infty
=
\frac{\log\beta}{-\epsilon}+O(1).
\label{eq:essential-scaling}
\end{equation}
Thus the deterministic order parameter vanishes with an essential singularity as $\alpha\to1^-$.

At the critical point $\alpha=1$, however, the behavior of a cascade initiated by a single instigator is governed by microscopic recruitment. Since
\begin{equation}
F(x)=1-(1-x)^\beta
=\beta x+O(x^2),
\end{equation}
the early recruitment process converges, as $N\to\infty$, to a Poisson Galton--Watson branching process with mean offspring number $\beta$.

This connection was established rigorously by Zeraoulia~\cite{zeraoulia_comment}. In particular, he showed that for $\alpha=1$ and $0<\beta<1$, the limiting cascade-size distribution is the Borel distribution
\begin{equation}
\lim_{N\to\infty}P_N(k)
=
\frac{e^{-\beta k}(\beta k)^{k-1}}{k!},
\qquad k=1,2,\ldots.
\label{eq:borel-critical}
\end{equation}
Because $\beta<1$, this branching process is subcritical, and its expected total progeny, including the initial instigator, is
\begin{equation}
\langle k\rangle\longrightarrow
\frac{1}{1-\beta}.
\label{eq:beta-progeny}
\end{equation}
Consequently,
\begin{equation}
\rho_\infty(N)\sim
\frac{1}{(1-\beta)N},
\qquad 0<\beta<1.
\label{eq:beta-critical}
\end{equation}
For $\beta=1/2$, the exact limiting amplitude is therefore $2$, rather than the value estimated by Monte Carlo simulations in Ref.~\cite{fontanari2026}.

We now extend this microscopic description to the finite-size critical region, allowing $\alpha$ to approach unity simultaneously with increasing
population size. This joint limit provides an  explicit scaling function that is not obtained from the critical Borel distribution alone.

Following the logarithmic scaling variable introduced in Ref.~\cite{fontanari2026}, we write
\begin{equation}
\alpha=1+\frac{z}{\log N},
\label{eq:log-window}
\end{equation}
where $z$ remains fixed as $N\to\infty$. Using Eq.~(\ref{eq:beta-small-x}), we find that for every fixed positive integer $j$,
\begin{equation}
NF(j/N)
=
\frac{N^{1-\alpha}j^\alpha}
{\alpha B(\alpha,\beta)}
[1+O(N^{-1})]
\longrightarrow \beta j e^{-z}.
\label{eq:local-recruitment}
\end{equation}
The factor $N^{1-\alpha}=e^{-z}$ therefore modifies the effective recruitment rate at microscopic cascade sizes.

In the joint scaling limit, the early recruitment process again converges to a Poisson Galton--Watson branching process, but now with effective mean offspring number
\begin{equation}
\lambda(z)=\beta e^{-z}.
\label{eq:effective-branching}
\end{equation}
Accordingly, the limiting distribution of every fixed cascade size is
\begin{equation}
\lim_{N\to\infty}P_N(k)
=
\frac{e^{-\lambda(z)k}
[\lambda(z)k]^{k-1}}{k!},
\qquad k=1,2,\ldots.
\label{eq:borel-scaling}
\end{equation}
When $\lambda(z)<1$, this is a normalized subcritical Borel distribution with finite mean
\begin{equation}
\langle k\rangle\longrightarrow
\frac{1}{1-\lambda(z)}
=
\frac{1}{1-\beta e^{-z}}.
\label{eq:log-scaling-function}
\end{equation}
Thus the finite-size scaling of the mean active fraction in the subcritical regime is
\begin{equation}
N\rho_\infty(N)
\longrightarrow
\mathcal{F}_\beta(z)
=
\frac{1}{1-\beta e^{-z}},
\qquad
z=(\alpha-1)\log N>\log\beta.
\label{eq:beta-log-scaling}
\end{equation}

Equation~(\ref{eq:beta-log-scaling}) is an explicit finite-size scaling prediction obtained by extending the microscopic branching-process description away from the critical line $\alpha=1$. It also provides a theoretical explanation for the logarithmic scaling variable previously identified from numerical data in Ref.~\cite{fontanari2026}.
At $z=0$, the scaling function reproduces Zeraoulia's critical amplitude,
\begin{equation}
\mathcal{F}_\beta(0)=\frac{1}{1-\beta},
\end{equation}
whereas for $z\to+\infty$, $\mathcal{F}_\beta(z)\to1$, corresponding to the isolated instigator.

The restriction $z>\log\beta$ has a direct physical interpretation. At $z=\log\beta$, the effective branching process becomes critical and its mean total progeny diverges. For $z<\log\beta$, the microscopic branching process is supercritical, and its finite-progeny distribution no longer accounts for the full
cascade-size distribution. In this regime, the contribution from cascades whose sizes grow with $N$ must also be considered.

The logarithmic scaling variable therefore describes the crossover between microscopic recruitment regimes, while Eq.~(\ref{eq:beta-log-scaling}) gives the explicit scaling function on its subcritical side. Together with the corrected macroscopic large-deviation rate derived in the present
work, this result provides a more complete description of the finite-size behavior near $\alpha_c=1$.


\section{Discussion}
\label{sec:discussion}

We have developed a systematic asymptotic analysis of the exact finite-population solution of Granovetter's threshold model~\cite{granovetter1978}. By expressing the propagation probability in terms of the empirical distribution of individual thresholds, we have reformulated the large-deviation problem as a constrained relative-entropy minimization. This approach avoids the limitations of the saddle-point approximation employed in Ref.~\cite{fontanari2026} and yields the correct macroscopic propagation rate for arbitrary continuous threshold distributions.

For Beta-distributed thresholds, the minimization problem admits an explicit solution characterized by free and pinned intervals of the optimal empirical threshold profile. Depending on the shape of the threshold density, four distinct geometries emerge: fully free, fully pinned, pinned--free, and free--pinned. The resulting analytical rate functions agree with finite-population importance-sampling estimates, while sequential Monte Carlo simulations independently reproduce the predicted conditional empirical profiles. Beyond determining the exponential probability of rare cascades, the formalism therefore identifies the threshold configurations responsible for their occurrence.

We have also revisited the critical behavior near $\alpha_c=1$ investigated in Ref.~\cite{fontanari2026}. Although that analysis employed an incorrect macroscopic propagation rate, the principal finite-size scaling laws remain valid. In particular, the mean final active fraction scales as $N^{-1/2}$ for uniform thresholds and as $N^{-1}$ for $\alpha=1$ and $0<\beta<1$, with the corrected critical amplitude $1/(1-\beta)$ established by Zeraoulia
\cite{zeraoulia_comment}. We have further derived an explicit scaling function for the logarithmic critical window $(\alpha-1)\log N$ in the subcritical microscopic regime, thereby connecting the finite-size behavior to the underlying branching-process description.

The main advantage of the present approach is its generality. The constrained-entropy formulation does not depend on the Beta family and can be applied to other threshold distributions, including those for which conventional saddle-point methods are difficult to justify. More broadly, the combination of empirical-measure large deviations and sequential activation constraints provides a framework for studying rare collective events in irreversible adoption, contagion, and cascade processes. Whenever the dynamics can be expressed through constraints on an empirical distribution, the same strategy may offer both the exponential rate of rare outcomes and a characterization of the microscopic configurations that produce them.

\section*{Acknowledgments}

JFF is partially supported by  Conselho Nacional de Desenvolvimento Cient\'{\i}fico e Tecnol\'ogico  grant number 300200/2026-9.

\appendix

\section{Technical justification of the finite-sample rate}
\label{app:technical}

In Section~\ref{sec:entropy}, we established a finite-sample criterion for the exponential propagation rate based on four properties of an auxiliary probability density $h_a$. For the optimal proposals $h_a=g_a^*$ constructed in Section~\ref{sec:beta}, Properties 1 and 3 follow directly from
their explicit form.

Here we verify the remaining two properties: the convergence of the likelihood grid averages (Property 2) and exponential
concentration of the likelihood weights (Property 4). These estimates are needed because the Beta density
\begin{equation}
f(t)=
\frac{t^{\alpha-1}(1-t)^{\beta-1}}
{B(\alpha,\beta)},
\qquad \alpha,\beta>0,
\end{equation}
may vanish or diverge at the origin.

Throughout this appendix, we consider sequences $a=n/N\to x$, with $x\in(0,1)$ fixed, and write
\begin{equation}
\ell_a(t)=\log\frac{q_a(t)}{h_a(t)},
\qquad
m_a=\int_0^a h_a(t)\ell_a(t)\,dt.
\end{equation}

\subsection{Convergence of the likelihood grid averages}

We first verify Property 2, namely
\begin{equation}
\frac1n\sum_{j=1}^n
\ell_a\left(\frac{aj}{n}\right)
=
\frac1a\int_0^a\ell_a(t)\,dt+o(1).
\label{eq:app-riemann}
\end{equation}
For the optimal Beta proposals, the log-likelihood ratio $\ell_a(t)$ is nondecreasing on $(0,a)$.

When $\alpha>1$, the Beta density vanishes at the origin as $f(t)\sim C t^{\alpha-1}$. On a pinned interval, where $h_a(t)=1/a$, we therefore have
\begin{equation}
\ell_a(t)
=
(\alpha-1)\log t+O(1),
\qquad t\to0.
\end{equation}
Although this expression diverges to minus infinity, its singularity is integrable. Moreover, the right-endpoint grid $a/n,2a/n,\ldots,a$ does not contain the singular point $t=0$.

Monotonicity gives the one-sided estimate
\begin{equation}
\frac1n\sum_{j=1}^{n-1}
\ell_a\left(\frac{aj}{n}\right)
\le
\frac1a\int_0^a\ell_a(t)\,dt.
\end{equation}
Adding the final grid value and using Property 3 yields
\begin{equation}
\frac1n\sum_{j=1}^n
\ell_a\left(\frac{aj}{n}\right)
\le
m_a+\frac{\ell_a(a)}{n}.
\label{eq:app-riemann-upper}
\end{equation}

For $a$ in a compact neighborhood of any fixed $x\in(0,1)$, $\ell_a(a)$ remains bounded. Thus, Eq.~(\ref{eq:app-riemann-upper}) already provides the one-sided estimate required for the upper bound in Section~\ref{subsec:bounds}.

To establish the full convergence in Eq.~(\ref{eq:app-riemann}), we separate the possible logarithmic singularity from the regular part of $\ell_a$.
The regular contributions are handled by ordinary Riemann-sum convergence. For the logarithmic contribution, the relevant sums reduce to expressions of the form
\begin{equation}
\sum_{j=1}^m\log\frac{j}{n}
=
\log(m!)-m\log n.
\end{equation}
Stirling's formula then shows that the difference between the normalized sum and its corresponding integral is $O(\log n/n)$.

The same estimate applies uniformly when $a$ approaches an interior point $x$, including the contact-transition points at which the pinned and free intervals change.
Consequently,
\begin{equation}
\frac1n\sum_{j=1}^n
\ell_a\left(\frac{aj}{n}\right)
=
m_a+O\left(\frac{\log n}{n}\right),
\end{equation}
which establishes Property 2.

\subsection{Exponential concentration of the likelihood weights}

We now verify Property 4. For every fixed $\varepsilon>0$, we must show that
\begin{equation}
\mathbb{P}_{h_a}
\left\{
\frac1n\sum_{i=1}^n
\ell_a(Y_i)<m_a-\varepsilon
\right\}
\le e^{-c_\varepsilon n},
\label{eq:app-concentration}
\end{equation}
for some $c_\varepsilon>0$, uniformly for $a$ sufficiently close to $x$. The variables $Y_1,\ldots,Y_n$ are independent samples from $h_a$.

For $\theta>0$, define
\begin{equation}
\Lambda_a(\theta)
=
\log\mathbb{E}_{h_a}
\left[e^{-\theta\ell_a(Y)}\right]
=
\log\int_0^a
h_a(t)e^{-\theta\ell_a(t)}\,dt.
\label{eq:app-lambda}
\end{equation}
Applying Markov's inequality to the positive random variable $\exp[-\theta\sum_i\ell_a(Y_i)]$ gives
\begin{equation}
\mathbb{P}_{h_a}
\left\{
\frac1n\sum_{i=1}^n
\ell_a(Y_i)<m_a-\varepsilon
\right\}
\quad\le
\exp\left\{
n\left[
\theta(m_a-\varepsilon)
+\Lambda_a(\theta)
\right]
\right\}.
\label{eq:app-chernoff}
\end{equation}

We must establish that the moment-generating function in Eq.~(\ref{eq:app-lambda}) is finite for sufficiently small positive $\theta$.

For nonincreasing Beta densities ($\alpha\le1$, $\beta\ge1$), the optimal proposal is $h_a=q_a$. Consequently, $\ell_a=0$, and the probability
in Eq.~(\ref{eq:app-concentration}) vanishes identically.

For U-shaped Beta densities ($\alpha<1$, $\beta<1$), the likelihood ratio is constant near the origin and bounded on the remainder of $(0,a)$. Thus, $\Lambda_a(\theta)$ is finite for every $\theta>0$.

For increasing and bell-shaped Beta densities ($\alpha>1$), the only possible unbounded negative contribution to $\ell_a(t)$ occurs on a pinned
interval near the origin, where
\begin{equation}
\ell_a(t)
=
(\alpha-1)\log t+O(1).
\end{equation}
Since $h_a(t)=1/a$ there, the integrand in Eq.~(\ref{eq:app-lambda}) behaves as
\begin{equation}
h_a(t)e^{-\theta\ell_a(t)}
\propto t^{-\theta(\alpha-1)}.
\end{equation}
This is integrable near $t=0$ whenever
\begin{equation}
0<\theta<\frac1{\alpha-1}.
\end{equation}

Hence, in every Beta regime, $\Lambda_a(\theta)$ is finite for sufficiently small positive $\theta$, uniformly for $a$ in a compact neighborhood of
any fixed $x\in(0,1)$. Furthermore,
\begin{equation}
\Lambda_a(0)=0,
\qquad
\Lambda_a'(0)
=
-\int_0^a h_a(t)\ell_a(t)\,dt
=-m_a.
\end{equation}
The local expansion is therefore
\begin{equation}
\Lambda_a(\theta)
=
-\theta m_a+O(\theta^2),
\end{equation}
with a remainder bounded uniformly for $a$ near $x$.
Substitution into Eq.~(\ref{eq:app-chernoff}) gives
\begin{equation}
\mathbb{P}_{h_a}
\left\{
\frac1n\sum_{i=1}^n
\ell_a(Y_i)<m_a-\varepsilon
\right\}
\le
\exp\left\{
-n\left[
\theta\varepsilon-O(\theta^2)
\right]
\right\}.
\end{equation}

For sufficiently small $\theta>0$, the coefficient $\theta\varepsilon-O(\theta^2)$ is strictly positive. Denoting it by $c_\varepsilon$ proves
Eq.~(\ref{eq:app-concentration}).

This completes the verification of Properties 2 and 4. Together with Properties 1 and 3 established for the optimal Beta proposals in Section~\ref{sec:beta}, these estimates justify the finite-sample large-deviation rate derived in Section~\ref{sec:entropy}.


\section{Optimality of the Beta profiles}
\label{app:beta-optimality}

We verify that the profiles constructed in Section~\ref{sec:beta} minimize relative entropy under the cumulative constraint.

Let $h$ denote one of the proposed optimal densities, with cumulative distribution $H(t)=\int_0^t h(s)\,ds$, and define
\begin{equation}
\varphi(t)=\log\frac{h(t)}{q_x(t)}.
\end{equation}
For any other admissible density $g$, with cumulative distribution $G$, the relative-entropy identity
\begin{equation}
D_{\mathrm{KL}}(g\Vert q_x)
-D_{\mathrm{KL}}(h\Vert q_x)
=
D_{\mathrm{KL}}(g\Vert h)
+\int_0^x[g(t)-h(t)]\varphi(t)\,dt
\label{eq:KLidentity}
\end{equation}
follows by expanding the logarithms. The first term on the right-hand side is nonnegative. We therefore need only show that the second term is also nonnegative.

The proposed Beta profiles have two types of intervals. On free intervals, $h$ is proportional to $q_x$, so that $\varphi$ is constant. On pinned
intervals, $h=1/x$, and hence
\begin{equation}
\varphi'(t)
=-\frac{f'(t)}{f(t)}\le0,
\end{equation}
because the original density is nondecreasing on every pinned interval.

Since both $g$ and $h$ are normalized, $G(0)-H(0)=G(x)-H(x)=0$. Integration by parts gives
\begin{equation}
\int_0^x(g-h)\varphi\,dt
=
-\int_0^x(G-H)\varphi'(t)\,dt.
\end{equation}
The boundary terms at internal junctions cancel because $\varphi$ is continuous. On free intervals, $\varphi'=0$. On pinned intervals, $H(t)=t/x$,
whereas admissibility requires $G(t)\ge t/x$. Therefore $G-H\ge0$ wherever $\varphi'$  may be nonzero. Since $\varphi'\le0$ on those intervals, we conclude that
\begin{equation}
\int_0^x(g-h)\varphi\,dt\ge0.
\end{equation}
Equation~(\ref{eq:KLidentity}) now implies
\begin{equation}
D_{\mathrm{KL}}(g\Vert q_x)
\ge D_{\mathrm{KL}}(h\Vert q_x).
\end{equation}
Thus, every proposed profile is a global entropy minimizer. Strict convexity of relative entropy establishes uniqueness up to sets of measure zero.

\subsection{Equality of the likelihood averages}

We also verify Property 3 of Section~\ref{sec:entropy}.
At the finite endpoint $a$, let $h_a$ be the optimal proposal and write
\begin{equation}
\ell_a(t)=\log\frac{q_a(t)}{h_a(t)}.
\end{equation}
On pinned intervals, $h_a=1/a$, so the spatial and statistical likelihood averages agree pointwise.
On a free interval $(u,v)$, the likelihood ratio $\ell_a$ is constant, say $\ell_0$. Moreover, the endpoints of the interval lie on the propagation boundary:
\begin{equation}
H_a(u)=\frac{u}{a},
\qquad
H_a(v)=\frac{v}{a}.
\end{equation}
These identities also hold when $u=0$ or $v=a$.
Consequently,
\begin{equation}
\int_u^v h_a(t)\,dt
=
\frac{v-u}{a},
\end{equation}
and therefore
\begin{equation}
\int_u^v h_a(t)\ell_a(t)\,dt
=
\ell_0\frac{v-u}{a}
=
\frac1a\int_u^v\ell_a(t)\,dt.
\end{equation}

Adding the contributions from all free and pinned intervals gives
\begin{equation}
\int_0^a h_a(t)\ell_a(t)\,dt
=
\frac1a\int_0^a\ell_a(t)\,dt.
\end{equation}
This proves Property 3. The equality depends on the piecewise structure of the optimal proposals and does not hold for a general change of measure.


\section{Matching points and feasibility of the Beta profiles}
\label{app:beta-geometry}

We establish the properties of the matching points introduced for bell-shaped and U-shaped Beta densities.

\subsection{Bell-shaped densities}

Suppose $\alpha>1$ and $\beta>1$, and let
\begin{equation}
m=\frac{\alpha-1}{\alpha+\beta-2}
\end{equation}
be the mode. For $x>m$, define
\begin{equation}
\Phi_x(c)
=
F(x)-F(c)-(x-c)f(c).
\end{equation}
The matching condition (\ref{eq:c-bell}) is equivalent to $\Phi_x(c)=0$. Since $f(0)=0$,
\begin{equation}
\Phi_x(0)=F(x)>0.
\end{equation}
Moreover,
\begin{equation}
\Phi_x(m)
=
\int_m^x[f(t)-f(m)]\,dt<0,
\end{equation}
because $f$ decreases strictly to the right of its mode. Differentiation gives
\begin{equation}
\Phi_x'(c)=-(x-c)f'(c).
\end{equation}
Since $f'(c)>0$ for $0<c<m$, the function $\Phi_x$ is strictly decreasing on this interval. It follows that there is exactly one root $c_x\in(0,m)$.

We next verify that the corresponding optimal profile satisfies the propagation constraint.

For $t\le c_x$, the profile is uniform, so
\begin{equation}
H(t)=\frac{t}{x}.
\end{equation}
For $c_x<t<x$, define
\begin{equation}
D(t)=H(t)-\frac{t}{x}.
\end{equation}
The matching condition implies
\begin{equation}
D(c_x)=D(x)=0.
\end{equation}
Furthermore,
\begin{equation}
D'(t)
=
\frac{f(t)-f(c_x)}{xf(c_x)}.
\end{equation}
Because $f$ is bell-shaped and $c_x<m$, the difference $f(t)-f(c_x)$ is initially positive and eventually negative as $t$ increases from $c_x$ to $x$. Thus, $D(t)$ first increases and then decreases to zero, without becoming negative. Consequently,
\begin{equation}
H(t)\ge\frac{t}{x}
\end{equation}
throughout the interval.

For $x\le m$, the fully pinned profile is trivially admissible.

\subsection{U-shaped densities}

Suppose $\alpha<1$ and $\beta<1$, and let
\begin{equation}
m=\frac{1-\alpha}{2-\alpha-\beta}
\end{equation}
be the minimum of the density. Define
\begin{equation}
\eta(t)=F(t)-tf(t).
\end{equation}
Its derivative is
\begin{equation}
\eta'(t)=-tf'(t).
\end{equation}
Since $f$ decreases on $(0,m)$ and increases on $(m,1)$, $\eta$ increases on $(0,m)$ and decreases on $(m,1)$.
Near the origin, $f(t)\sim Ct^{\alpha-1}$, and therefore
\begin{equation}
\eta(t)
\sim
\left(\frac1\alpha-1\right)tf(t)>0.
\end{equation}
As $t\to1$, the density diverges while $F(t)\to1$, so that
\begin{equation}
\eta(t)\to-\infty.
\end{equation}
There is consequently a unique $d\in(m,1)$ satisfying
\begin{equation}
\eta(d)=0,
\qquad F(d)=df(d).
\end{equation}
Now
\begin{equation}
\frac{d}{dt}\frac{F(t)}{t}
=
-\frac{\eta(t)}{t^2}.
\end{equation}
Hence $F(t)/t$ decreases on $(0,d)$ and increases on $(d,1)$. For $x\le d$, this implies
\begin{equation}
\frac{F(t)}{F(x)}
\ge\frac{t}{x},
\qquad 0<t\le x.
\end{equation}
Thus, the conditioned density $q_x$ is admissible.

For $x>d$, consider the piecewise profile (\ref{eq:g-U}). On $0<t\le d$, its cumulative distribution is
\begin{equation}
H(t)=\frac{F(t)}{xf(d)}.
\end{equation}
Since $F(t)/t\ge F(d)/d=f(d)$ for $t\le d$, we obtain
\begin{equation}
H(t)\ge\frac{t}{x}.
\end{equation}
At $t=d$, equality holds because $F(d)=df(d)$.
For $d<t<x$, the density equals $1/x$, so the cumulative distribution continues exactly along the propagation boundary:
\begin{equation}
H(t)=\frac{t}{x}.
\end{equation}
The proposed U-shaped profile is therefore admissible for every $x>d$.

Together with the optimality argument of Appendix~\ref{app:beta-optimality}, these results establish the four profiles used in Section~\ref{sec:beta}.



\end{document}